\documentclass[aps,prab,reprint,superscriptaddress]{revtex4-2}

\usepackage{dcolumn}
\usepackage{graphicx}
\usepackage{subfigure}
\usepackage{textcomp}
\usepackage{float}
\usepackage{amsmath}
\graphicspath{{figures/}}

\begin{document}
	
	
	\title{Longitudinal Compression of Macro Relativistic Electron Beam}
	
	
	\author{An Li}
	\affiliation{Department of Engineering Physics, Tsinghua University, Beijing 100084, People’s Republic of China}
	
	\author{Jiaru Shi}
	\affiliation{Department of Engineering Physics, Tsinghua University, Beijing 100084, People’s Republic of China}

	\author{Hao Zha}
	\email{ZhaH@mail.tsinghua.edu.cn}
	\affiliation{Department of Engineering Physics, Tsinghua University, Beijing 100084, People’s Republic of China}
	
	\author{Qiang Gao}
	\affiliation{Department of Engineering Physics, Tsinghua University, Beijing 100084, People’s Republic of China}
	
	\author{Liuyuan Zhou}
	\affiliation{Department of Engineering Physics, Tsinghua University, Beijing 100084, People’s Republic of China}
	
	\author{Huaibi Chen}
	\affiliation{Department of Engineering Physics, Tsinghua University, Beijing 100084, People’s Republic of China}
	
	
	\date{\today}
	
	\begin{abstract}
		
		We presented a novel concept of longitudinal bunch train compression capable of manipulating relativistic electron beam in range of hundreds of meters. This concept has the potential to compress the electron beam generated by conditional linear accelerator with a high ratio and raise its power to an high level comparable with large induction accelerators. The method utilizes the spiral motion of electrons in a uniform magnetic field to fold hundreds-of-meters-long trajectories into a compact set-up. The interval between bunches can be adjusted by modulating their sprial movement. The method is explored with particle dynamic simulation. Compared to set-up of similar size, such as chicane, this method can compress bunches at distinct larger scales, opening up new possibilities generating beam of high power with compact devices and at lower costs.

	\end{abstract}
	
	
	\maketitle
	

	\section{introduction}
	
	Ultrahigh pulsed power supply has remained a bottleneck in numerous frontier studies in recent years, including fuel pellet compression and ignition in fusion research\cite{FastIgnition,ZPinch,ECRH}, high power wake-field stimulation for novel accelerators\cite{WakefieldAcc,WakefieldAcc2,WakefieldAcc3}, and ultrahigh dose rate X-Ray generation for FLASH radiotherapy or FLASH radiography applications \cite{FLASHRadiotherapy0,FLASHRadiotherapy,DAHRT,FLASHRadiography}.

	To generate pulsed power, energy could be stored in mediums such as laser\cite{CPA,LaserCompression}, electric current\cite{SolidModulator,SolidModulator2}, microwave\cite{MicrowaveModulator,MicrowaveModulator2}, or electron beam\cite{MicrowaveModulator,MicrowaveModulator2}, then get released in short pulse duration. Among these options, electron beam has certain advantages, such as high energy conversion efficiency from AC power, and negligible energy dissipation during compression. Furthermore, as space chage effect decreases with increased particle energy $E$, the power capacity of electron beam is proportional to $E^3$. This indicates an ultrahigh energy storage limit for relativistic beam.
	
	Nevertheless, directly accelerating an high current electron beam to the required energy is both technically challenging and expensive. Induction accelerators used in scientific installations for Z-Pinch or FLASH radiography are typically hundreds of meters long\cite{CLIC2,DAHRT,ZPinch2}, and require a large number of klystrons for power supply. It's more economical to accelerate a long pulse beam with an applicable current, and then compress it to a short duration for power multiplication.

	Electron beam compression schemes that operate at femtosecond or picosecond timescales have been extensively investigated lately\cite{VelocityBunching,MagnetCompress,LaserCompress,CoulombCompress}. While compression for macro beam in nanoseconds to microseconds duration is less mentioned. For relativistic particle, the trajectory span hundreds of meters in spatial scale, resulting in high costs and large set-up dimensions. For example, the Compact Linear Collider (CLIC) study\cite{CLIC2} proposed a bunch train combination system with combiner ring circumference of up to 438 meters to manipulates beam of several microseconds. And the Geel Electron LINear Accelerator Facility(GELINA)\cite{GELINA,GELINA2} employed a magnetic compression system weights 50 tons to compress beam with pulse width of 10 nanoseconds.

	Regarding this issue, we proposed a method for compressing electron bunch train that capable of manipulating beam of tens of nanoseconds with a relatively compact set-up, providing new possibilities for the low-cost production of the high-current relativistic electron beams.

\begin{figure}[htbp]
		\includegraphics[width=8.5cm]{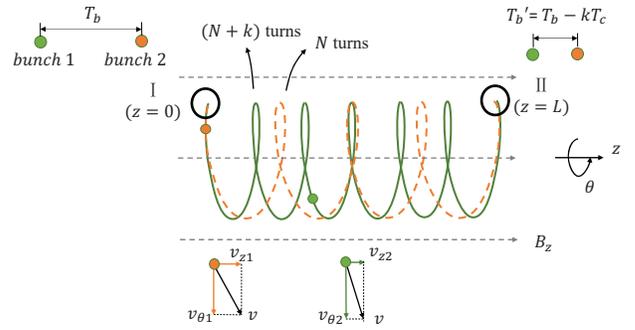}
		\caption{Compressing principle. After deflected by the time-varying magnet, the axial velocty and the spiral pitch of the spiral motion of each bunch get different. }
		\label{FIG.1}
	\end{figure}

	As illustrated in FIG.\ref{FIG.1}, The spiral motion of electrons in a uniform magnetic field can be harnessed to fold hundreds of meters worth of trajectories into a compact cylindrical volume. This process allows for the efficient use of space, making it an attractive option for the electron bunch train compression. By modulating the spiral helix of a train of electron bunches, we are able to manipulate the intervals between the bunches and achieve beam compression.
	
	This article presents a detailed introduction to the compression device, and we demonstrate its ability to manipulate the particle beam in two schemes, namely, bunch train compression and bunch train combination, with particle dynimic simulation.

	\section{Installation}
	
	To generate a uniform axial magnetic field within a cylindrical cavity, we utilized a large solenoid coil as the primary component of the compression device. Furthermore, we incorporated a combination of permanent magnets as the injection and extraction ports and positioned electromagnets within the cavity to modulate the spiral motion of the train of electron bunches.
	
	\begin{figure}[htbp]
		\includegraphics[width=8.5cm]{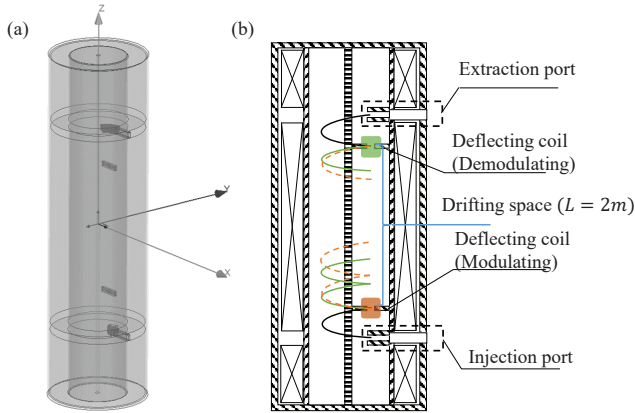}
		\caption{(a) Compression installation (b) The front view Profile of the compression installation. The setup comprises a primary solenoid, injection and extraction ports, and several deflecting magnets inside the cavity.}
		\label{FIG.2}
	\end{figure}

		\subsection{Solenoid}
		The solenoid coil used in our compression device has a length of 5 meters, an outer diameter of 1 meter, and an inner diameter of 0.8 meters. The injection and extraction ports are separated by an axial distance of 3 meters. The coil is divided into three sections, each of which has independently adjustable current, enabling precise control of the magnetic field distribution within the cylindrical cavity.
		
\begin{figure}[htbp]
			\includegraphics[width=8.5cm]{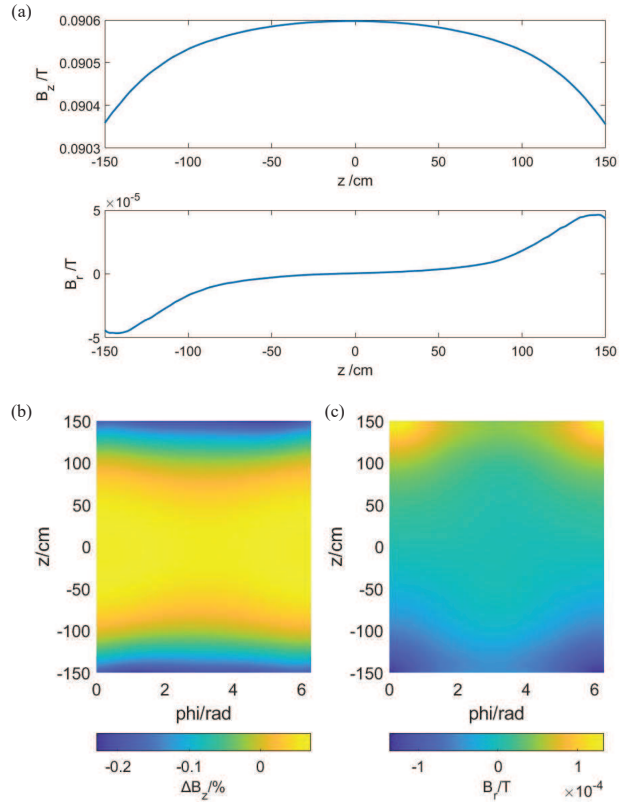}
			\caption{Uniformilty of magnetic field in the cavity. (a) Magnetic field component $B_z$ and $B_r$ along the z-direction at $R = 25 \; cm$ (b) Uniformity of the axial magnetic field $B_z$ on cylindrical surface $R=25 \;cm$, where $\Delta B_z = \frac{B_z-\overline{B_z}}{\overline{B_z}}$ (c) Distribution of radial magnetic field $B_r$}
			\label{FIG.3}
		\end{figure}

		In the design, the average magnetic field strength within the cavity is approximately 0.09 T. As electron beam of 6.5 MeV is employed in simulation, the reference spiral radius is 25 cm.  The uniformity of the magnetic field on the cylinderical surface of $R = 25$ cm is shown in FIG. \ref{FIG.3}. The axial magnetic field $B_z$ has a maximum non-uniformity of less than 0.3\%, and the radial component $B_r$ is below $10^{-4}$ T.
		
		\subsection{Injection and extraction ports}
		The purpose of the injection structure is to induce spiral motion of the electron beam along the axis of the cylindrical cavity after injection. It is composed of ferromagnetic shells and permanent magnets with reversed magnetization to the uniform field.
		
\begin{figure}[htbp]
			\includegraphics[width=8.5cm]{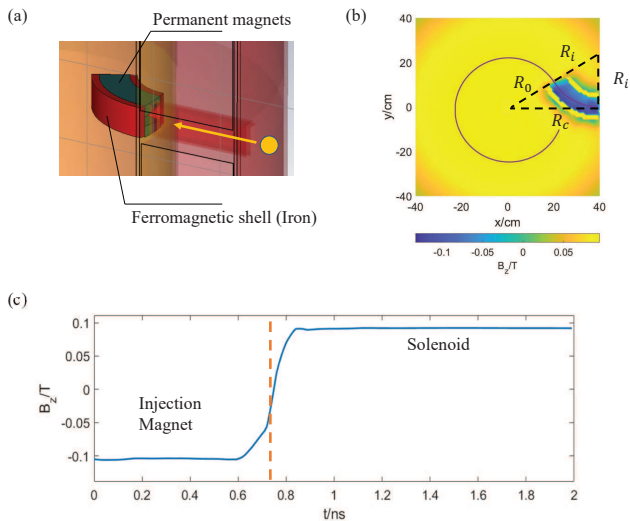}
			\caption{(a) Structure of injection port (b) Magnetic field $B_z$ on injection plane and the simulated injected electron trajectory. (c) $B_z$ observed on injected electron's trajectory.}
			\label{FIG.4}
		\end{figure}

		The projection of the trajectory of the injected electrons consists of two arcs, as shown in FIG. \ref{FIG.4}(b). $R_i$ and $R_0$ represent the radii of the electron's circular motion at the injection port and within the cavity, respectively, while $R_c$ is the radius of the cylindrical cavity. With $R_0 = 25 $ cm, $R_c = 40$ cm, and the geometric relationship, we can calculate that $R_i = 19.5$ cm, which corresponds to a magnetic field of $B_i = -0.11 $ T for electrons with an energy of $6.5$ MeV.
		
		The magnetic field distribution observed along the path of the injected electrons is shown in FIG. \ref{FIG.4}(c).
		
		The extraction structure is symmetrically opposite to the injection structure to facilitate the extraction of spiral bunches from the solenoid. 
		
		\subsection{Deflecting magnets}
		
		The deflecting magnets play a crucial role in the electron bunch train compression method. Positioned along the trajectory of the introduced electrons, they generate a time-varying, radial-oriented, dipole magnetic field that deflects the incoming beam and alters bunches' axial velocity by a value of $\Delta v_z$. The time-varying deflecting effect modulates the helical trajectory of the bunch train, manipulating the interval between bunches and ensuring successful extraction of each bunch.
		
\begin{figure}[htbp]
			\includegraphics[width=8.5cm]{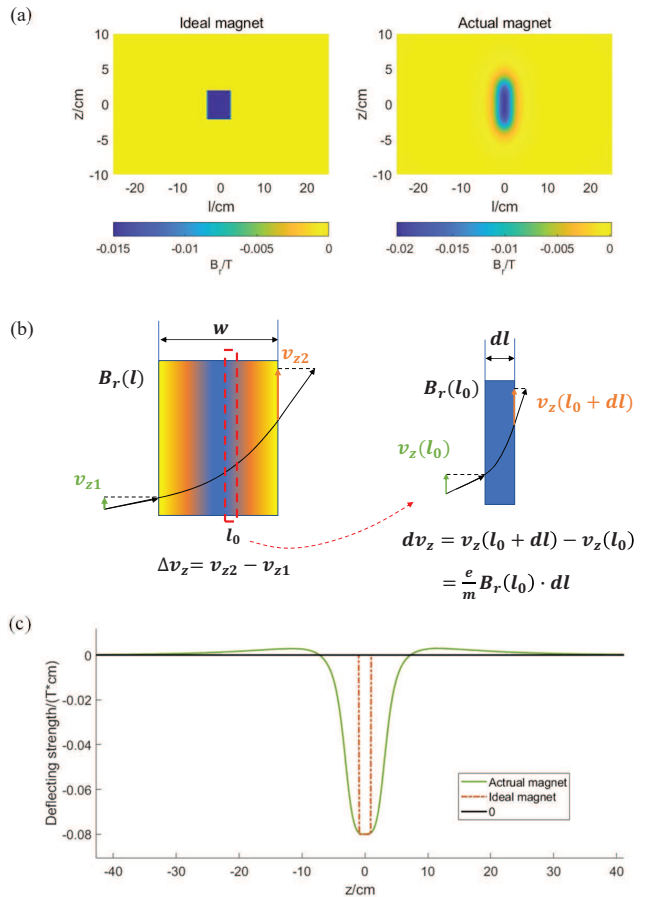}
			\caption{(a)  Magnetic field of ideal and actual deflecting magnets in cylindrical surface of $R = 25 cm$  (b) Deflecting strength $\Delta v$ of the non-uniform magnetic field (c) Integral field $\int B(l)\cdot dl$ of ideal and actual defelcting magnets}
			\label{FIG.5}
		\end{figure}

		It should be noted that the deflecting magnets are composed of coils without a magnetic yoke, as the yoke can cause distortion in the uniform axial field and the beam's stable spiral movement. Therefore, compared to the ideal situation of a sharp boundary magnetic field, the deflecting magnets have non-negligible fringe fields. The magnetic field distribution of the magnet on the cylindrical surface of R = 25 cm can be seen in FIG. \ref{FIG.5}(a).
		
		For this non-uniformly distributed magnetic field, its deflecting effect on electron motion is shown in FIG. \ref{FIG.5}(b). Taking a narrow slice of width $dl$ at $l = l_0$, the magnetic flux in the slice is approximately uniform ($B_r = B_r(l_0)$), and the deflecting effect of the slice on the electron is $dv_z = \frac{e}{m}B_r(l_0)\cdot dl$, where $e$ and $m$ is charge and mass of the relativistic electron. Therefore, the deflecting strength of the magnet can be expressed as $\Delta v_z = \int \frac{e}{m}B_r(l)\cdot dl$.
		
		The distribution of the Integral field $\int B(l)\cdot dl$ of the deflecting magnet along the z direction can be seen in FIG. \ref{FIG.5}(c).The actual deflecting magnet displays a wider range of magnetic field distribution when compared to the ideal one. The magnet's fringe field extends along the axial direction (z-direction) for almost 60 cm, which is exceptionally greater than the designed good field region width of 2 cm.
		
		The fringe field effect significantly interferes with our magnet and system design. Due to the fringe field, electrons located outside the magnet's good field region are still subject to the magnet's deflection, especially when the magnet's current varies with time. This can result in two outcomes: firstly, the trajectory of the injected electron beam may deviate from the magnet's good field region. Secondly, electrons may remain affected by the fringe field after being deflected by the magnet, leading to a non-uniform pitch of the spiral motion.

		In the following section, we will provide a detailed explanation of our compression method and how we aim to effectively address the fringe field issue.

	\section{Compressing methods}
	
	Based on the aforementioned installation, we have developed two distinct compression modes: bunch train compression and bunch train combination. These modes differ slightly in both principle and device design. We will discuss the fundamental principles of these two compression methods and present the specific compression results achieved in simulation.
	
		\subsection{Bunch train compression}
		
		In the bunch train compression mode, our goal is to decrease the interval between adjacent bunches in the train while increasing the average current and power of the beam with a ratio of $\eta = \frac{T_b}{T_b'}$.
		
		\begin{figure}[htbp]
			\includegraphics[width=8.5cm]{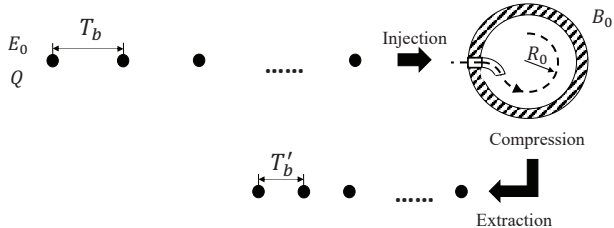}
			\caption{Sketch of bunch train compression. }
			\label{FIG.6}
			
		\end{figure}

			\subsubsection{Compressing principle}
			
			To illustrate, let us consider the compression of two electron bunches depicted in FIG. \ref{FIG.1}. Assuming that the time interval between the injected bunch 1 and bunch 2 is $T_b$, and the cyclotron period of the bunch in the uniform magnetic field $B_z$ is $T_c$. After being modulated by the deflecting magnet, bunch 1 and bunch 2 spiral for $N$ and $N-k$ turns before extraction, respectively.  As a result, the spacing between the two bunches is reduced to $T_b'=T_b-kT_c$ after extraction, and a compression ratio $\eta = \frac{T_b}{|T_b-kT_c|}$ is achieved. The beam is compressed when $|\eta |>1$. If $|\eta |<1$, the beam is diluted.

			\begin{figure*}[htbp]
				\includegraphics[width=13cm]{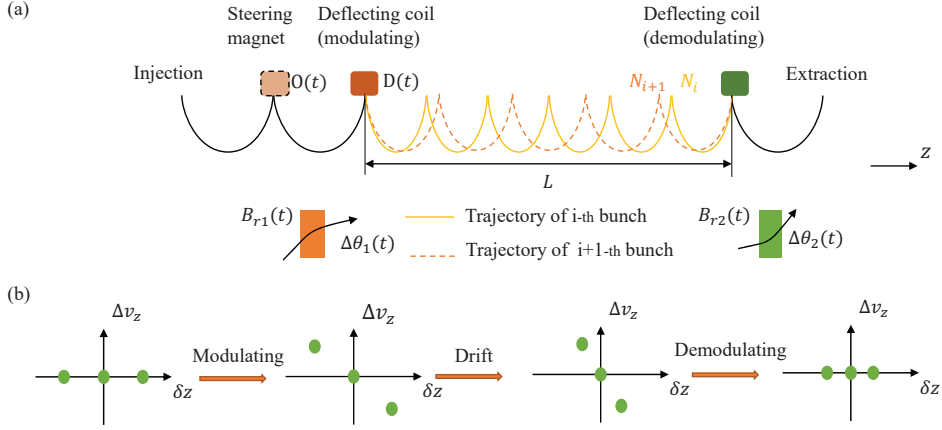}
				\caption{(a)Sketch of the compression system. (b) Compression principle presented by phase space distribution of bunch train in z-direction}
				\label{FIG.7}
			\end{figure*}

			This compression method is similar to velocity bunching, but tailored for relativistic electron beams. It is impossible to create a considerable velocity difference between the head and tail of a relativistic electron beam through energy modulation. However, by utilizing the helical motion of electrons in a uniform magnetic field, we can expand the electrons' trajectories from one dimension to three dimensions. By modulating the axial component of the electrons' velocity, we can achieve a "velocity bunching" method skillfully for the relativistic electron beams.

			FIG. \ref{FIG.7}(a) depicts a schematic diagram of the bunch train compression system. In this mode, the structure and field distribution of the modulating magnet and demodulating magnet are consistent with the deflecting magnet described earlier, and they produce time-varying deflection effects on the incoming bunch train. FIG. \ref{FIG.7}(b) illustrates the modulation and demodulation of the bunches in the axial phase space. 
			
			Assuming the fringe field problem is neglected, and the bunch train is injected with an axial velocity of $v_{z0}$, the modulating magnet should deflect the i-th bunch by $\Delta v_i = \frac{L}{N_i} - v_{z0}$, where $L$ is the axial length of the drift space and $N_i$ is the number of spiral turns of the i-th bunch in the drift space. This ensures all the bunches to get acess at the demodulating magnet.Similarly, the demodulating magnet should deflect the bunches by $\Delta v_i' = v_{z0} - \frac{L}{N_i}$, successively recovering the axial velocity of the bunches and preparing them for extraction.
			
			However, due to the existence of fringe fields, the magnets will deflect the motion of all bunches within the fringe field range at every moment, greatly increasing the complexity of the problem.

			To address this issue, a steering magnet was added between the injection port and modulating magnet, as shown in FIG. \ref{FIG.7}(a). This magnet, has similar sturcture of the deflecting magnet, produces a time-varying radial magnetic field to counteract the effect of the modulating magnet's fringe field on the injection side. An iterative calculation method was employed to obtain the time-varying currents of the three magnets. The objective of the calculation was to ensure that each bunch reaches the center of the modulating magnet's good field region after injection, passes through a specified number of spiral movements, reaches the center of the demodulating magnet's good field region, and finally exits the solenoid installation through the extraction port. After determining the geometry of the magnets and the parameters of the bunch train, we can solve for the required time-varying strength of each magnet.
			
\begin{figure}[htbp]
				\includegraphics[width=6cm]{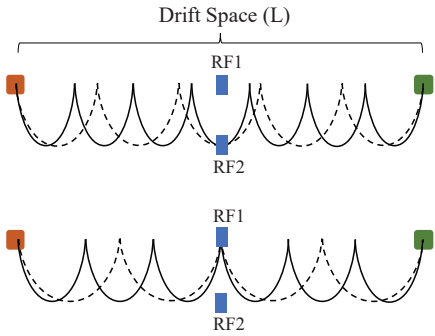}
				\caption{Position of RF electrodes. }
				\label{FIG.8}
			\end{figure}

			In addition, as the turning period $T_c$ is proportional to electron's energy, dispersion of the beam can cause the longitudinal size of the bunches to grow, resulting in a deviation of the electrons' arrival time on the demodulating magnet. This time difference can cause incorrect demodulation of the electron's axial velocity and lead to particle loss. Since the trajectory of each bunch is different, it is difficult to find suitable positions to place components for control and focusing. After comparing a series of options, we finally chose to place two RF electrodes in the middle of the dirft space for focusing purpose, as shown in FIG. \ref{FIG.8}. Trajectories of Bunches with even $N$ intersect at the RF1 Field, while the odd $N$ trajectories intersect at the RF2 Field. The RF Fields can provide a certain degree of longitudinal focusing on each bunch. However, due to the difference in drift time, the focusing effect on each bunch is not the same. In our design, the voltage of RF electrodes are chosen for minimizing the average size of extracted bunches.

			\subsubsection{simulation results}
			
			According to FIG. \ref{FIG.6}, our simulation assumed a solenoid cavity with a uniform magnetic field strength of $B_z = 0.09$ T, an electron beam with an average energy of $E_0 = 6.5 $ MeV, and each bunch having a charge of $Q = 1$ nC.  The cyclotron radius of the electrons in the uniform magnetic field is $R_0 = 25  $ cm and the period is $T_c =  5.4$ ns.  We employ a beam consisting of 10 bunches with an initial interval of $T_b = 6$ ns, corresponding to a beam length of approximately 60 ns.

			During the compression process, the first bunch undergoes 17 spiral periods from injection to the extraction port, the second bunch takes 16, and so on. 
			
\begin{figure}[htbp]
				\includegraphics[width=8.5cm]{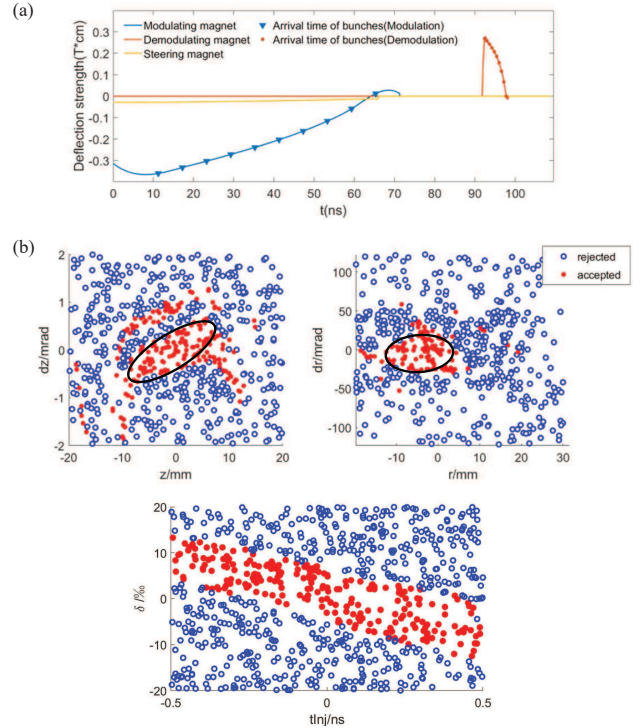}
				\caption{(a) Time varying deflecting strength of deflecting mangets (b) Phase space acceptance of the compression system in vertical $<z>$, horizontal $<r>$,  and longitudinal direction. $\delta = \delta E/E_0$. }
				\label{FIG.9}
			\end{figure}

			The time-varying deflection strengths of the three electromagnetic coils obtained through iterative calculation are shown in FIG. \ref{FIG.9}(a). The amplitude of the RF electrode used for longitudinal focusing is 250 kV, and the frequency is 185 MHz.
			
			The phase space acceptance  of this compression system is shown in FIG. \ref{FIG.9}(b). The geometric acceptance in $z$ and $r$ directions are $15 \; mm \cdot mrad$ and  $780 \; mm \cdot mrad$. The aperture of this system is quite large. The energy acceptance range of the system for 6.5 MeV beam is approximatly $\pm 5 \text{\textperthousand}$. If each bunch with dispersion $|\delta| <  5 \text{\textperthousand}$ is injected with a pulse width of less than 0.2 ns, particle losses caused by deflection errors during the modulating and demodulatng process can be completely avoided, thus achieving 100\% compression efficiency.
			
\begin{figure}[htbp]
				\includegraphics[width=8.5cm]{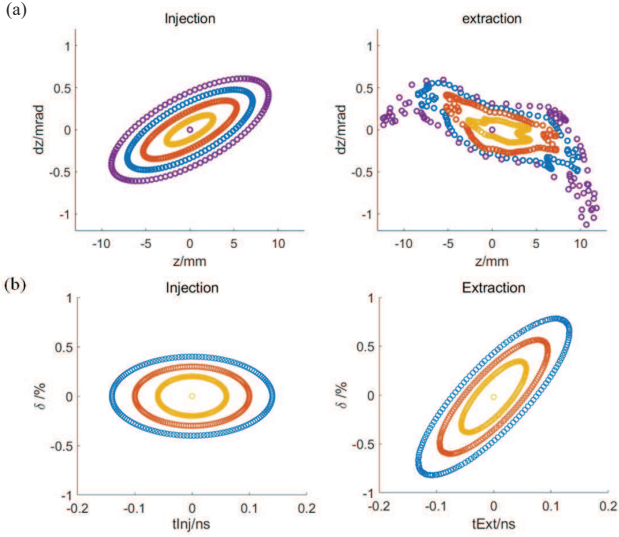}
				\caption{Phase-space distribution transformation of the first injected bunch before and after the compression in (a)$z$ direction and (b) longitudinal direction}
				\label{FIG.10}
			\end{figure}

			The phase-space distribution transformation of the first injected bunch before and after the compression in $z$ and longitudinal direction is shown in FIG. \ref{FIG.10}. the magnet's fringe field introduces nonlinearity in this transmission system, causing some distortion to bunch's transverse ellipse envelope.
			
\begin{figure}[htbp]
				\includegraphics[width=8.5cm]{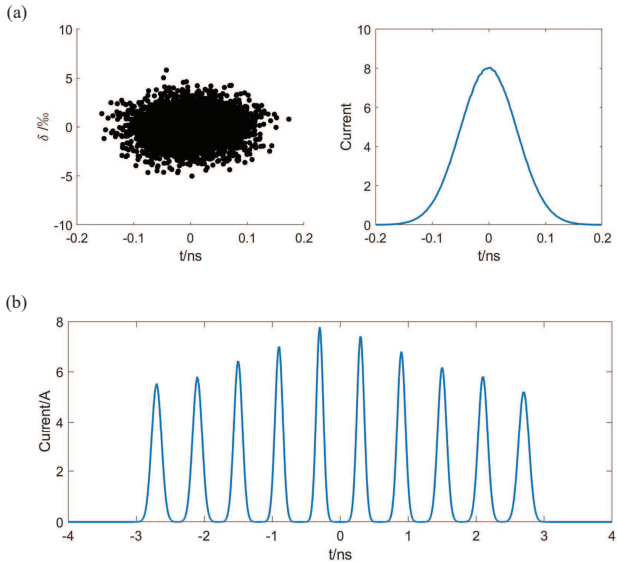}
				\caption{(a)Profile of each injected bunch (b)Current profile of compressed bunch train}
				\label{FIG.11}
			\end{figure}

			In the simulation, the profile of each bunch in the bunch train before injection is shown in FIG. \ref{FIG.11}(a). The bunch has a initial RMS length of 0.1 ns and maximum energy spread of $\delta = \pm 5 \text{\textperthousand}$. 
			
			After compression, current profile of the bunch train is illustrated in FIG. \ref{FIG.11}(b). The HWFM of each bunch is different because of the inconsistant focusing strength of the RF electrodes. 
			
			The interval between the bunches is reduced from $T_b = 6 $ ns to $T_b' = T_b - T_c = 0.6$ ns, resulting in a compression with ratio of $\eta = 10$. The average current of the beam string increased from 0.17 A to  1.7 A, and the average power increased from 1.1 MW to 11 MW.

			We have identified several limiting factors for further improving the compression ratio. The magnet's fringe field limits the minimum pitch of the electron spiral trajectory, which restricts the maximum number of bunches in the bunch train. Moreover, the time-varying voltage of the demodulation magnet limits the minimum spacing between the extracted bunches,  particularly when a magnet without a yoke is used. If $T_b'$ is further reduced, a faster change in demodulating strength is required. However, there is a limiting case, $T_b' = 0$, that can be achieved by employing another demodulation method.
			
		\subsection{Bunch train combination}
		
			\subsubsection{Combining principle}
			
			If the injected bunch interval $T_b$ is equal to cyclotron period $T_c$, all the bunches will arrive at the demodulating magnet simultaneously after the spiral procedure. In this case, the time-varying magnet described in the previous section cannot demodulate the axial velociy of the bunch train.
			
			We can employ a spatial-gradient magnet instead of the time-gradient magnet for demodulation, and focus all the bunches at the extraction port. 
			
\begin{figure}[htbp]
				\includegraphics[width=8.5cm]{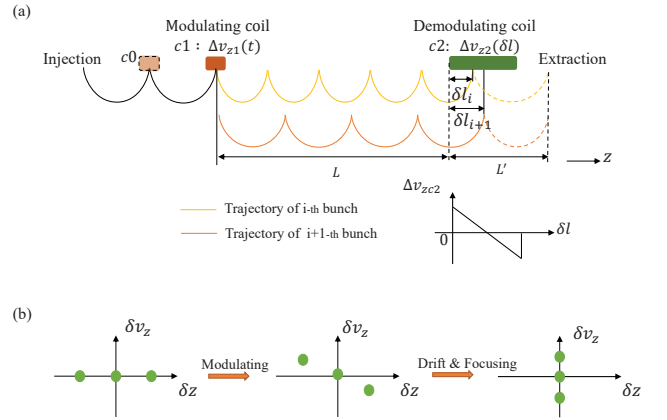}
				\caption{(a) Sketch of the combination system. (b) Combination principle presented by phase space distribution of bunch train in z-direction}
				\label{FIG.12}
			\end{figure}

			The combination system is illustrated in FIG. \ref{FIG.12}(a). The ideal deflecting strength of demodulating magnet is linearly related to the coordinate $\delta l$. When its strength distribution along the $z$ direction $\Delta v_{z2}(\delta l)$ is determined, the motion of the i-th bunch in the axial direction should satisfy the following equations.

			$$
			\begin{aligned}
				(v_{z0} + \Delta v_{z1}(t_i)) \cdot N_iT_c &= L+\delta l_i \\
				(v_{z0} + \Delta{v_{z1}(t_i)} + \Delta v_{z2}(\delta l_i)) \cdot T_c &= L'-\delta l_i\\
			\end{aligned}
			$$
			
			From a phase space perspective, as shown in FIG. \ref{FIG.12}(b), we rotate the bunch train that originally distributed along the z-axis to the $\delta v_z$-axis after the spiral procedure. At extraction port, from the spatial perspective, bunches overlap with each other and seem like a large bunch.
			
\begin{figure}[htbp]
				\includegraphics[width=8.5cm]{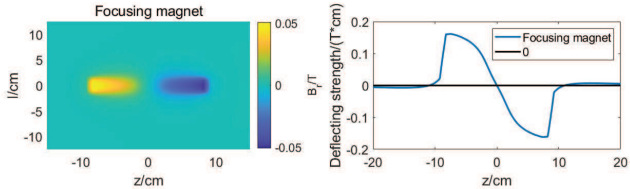}
				\caption{Radial magnetic flux and deflecting strength distribution of demodulating magnets }
				\label{FIG.13}
			\end{figure}

			Magnetic flux density on cylindrical surface $R = 25 \; cm$ and deflecting strength $D(z) = \int B(l,z) dl$ along the z direction of the demodulating magnet are shown in FIG. \ref{FIG.13}. The range of the good field region that satisfies the linear relationship is approximately 6 cm.
			
			In the presence of fringe fields, by iteratively solving for the time-varying strength of c0 and c1, as well as the value of $\delta l_i$ for each bunch, the magnet parameters applicable to bunch train combination can be obtained.

			\subsubsection{simulation results}
			In the simulation of bunch train combination, parameters of solenoid field, injection and extraction port are consistent with the compression mode. Combination of 10 bunches with charge of 1 nC and initial interval $T_b = T_c$ is performed. 
			
			\begin{figure}[htbp]
				\includegraphics[width=8.5cm]{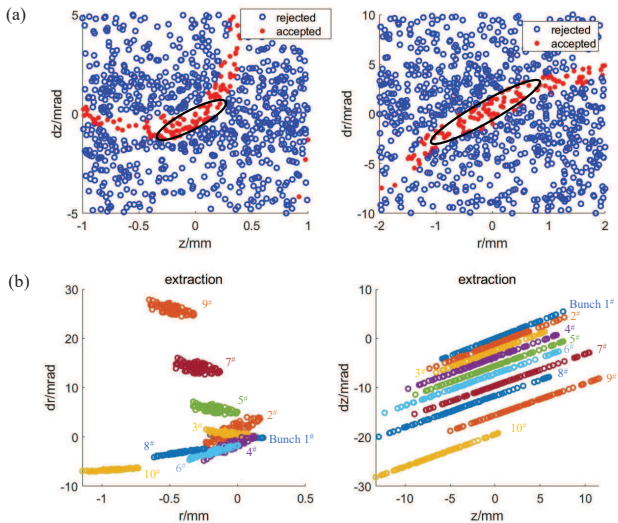}
				\caption{(a)Phase space acceptance of the combination system. (b) Phase space distribution of conbined bunch.}
				\label{FIG.14}
			\end{figure}
		
			The phase-space acceptance of the combinaiton system is shown in FIG. \ref{FIG.14}(a).The geometric acceptance in $z$ and $r$ directions are $1.2 \; mm \cdot mrad$ and  $2.5 \; mm \cdot mrad$.
	
			Before injection, each bunch has an initial emittance of $1 \; mm \cdot mrad$ and, after the spiral procedure, the phase space distribution of the bunch train at the extraction port is stacked along the z' direction to form a large bunch with a charge of 10 nC (FIG. \ref{FIG.14}(b)).

			While this combination method produces bunch with a large charge, the emittance is also large. If the bunch keeps on drifting for a period of time without post-focusing, it will spread out again. Therefore, this method may not be suitable for applications that are strict on electron beam emittance, while it is appropriate for applications such as white spectrum neutron sources, where the combined bunch hits the target after combination and the emittance of the bunch is insignificant.

	\section{Summary and Outlook}
	
	In this letter, we demonstrate a macro electron beam compression concept achieved by modulating bunches’ spiral motion in a uniform magnetic field. With proposed set-up, simulation of 2 different compressing schemes are performed.
	
	Our method is capable of compressing electron beam on a scale of tens of nanoseconds using a relatively compact device (<10 meters in spatial scale). It offers high compression efficiency and large energy storage limits, making it a promising choice for power compression.
	
	We will conduct further research on this compression method, attempting to optimize it and achieve electron beam compression in experiments.

	\bibliography{ref}
	
\end{document}